\documentclass{llncs}
\usepackage{amsmath}
\usepackage{amssymb}
\usepackage{epsfig}
\usepackage{subfigure}
\usepackage{epic}
\usepackage[boxed]{algorithm2e}

\def\NC{\newcommand}
\def\rNC{\renewcommand}

\rNC{\leq}               {\leqslant}
\rNC{\geq}               {\geqslant}
\rNC{\epsilon}           {\varepsilon}

\NC{\eg}                 {e.g.}
\NC{\ie}                 {i.e.}
\NC{\todo}            [1]{{\bf TODO:}\marginpar{{\bf [!!!]}} #1}
\NC{\rant}            [1]{\marginpar{{\bf [!!!]}}{\em [#1]}}

\NC{\chebychev}          {Chebychev}

\NC{\bigo}            [1]{O(#1)}
\NC{\reals}              {\mathbb{R}}
\NC{\plane}              {\reals^2}
\NC{\norm}            [1]{\left\|#1\right\|}
\NC{\logn}               {\log n}
\NC{\lognsqr}            {\log^2 n}
\NC{\classP}             {P}
\NC{\nil}                {\varnothing}
\NC{\card}            [1]{\left|#1\right|}
\NC{\set}             [1]{\left\{#1\right\}}

\NC{\muest}              {\mu_{\mathrm{est}}}

\NC{\inV}                {\in V}
\NC{\inE}                {\in E}
\NC{\vinV}               {v\inV}
\NC{\einE}               {e\inE}
\NC{\posof}           [1]{x_{#1}}
\NC{\posofv}             {\posof{v}}
\NC{\boundarynodes}      {D}
\NC{\nodesinarea}     [1]{V(#1)}

\NC{\commradius}         {R}

\NC{\prob}            [1]{\Pr\!\left[#1\right]}
\NC{\binomial}        [2]{\mathrm{bin}\left(#1,#2\right)}
\NC{\expected}        [1]{\mathrm{E}\!\left[#1\right]}
\NC{\given}              {\;\pmb{|}\;}

\NC{\vol}             [1]{\lambda(#1)}
\NC{\unitvol}            {\lambda_{\circ}}
\NC{\netvol}             {\vol{\netarea}}

\NC{\labeledarea}     [1]{A_{#1}}
\NC{\netarea}            {\labeledarea{\mathrm{net}}}
\NC{\netboundary}        {\partial\netarea}
\NC{\ball}            [2]{B_{#1}(#2)}
\NC{\netball}         [2]{\labeledarea{#1}\!\left(#2\right)}

\NC{\average}         [1]{\mathrm{avg}(#1)}

\NC{\messagen}        [1]{{\sf #1}}
\NC{\messagel}        [2]{\messagen{#1}{\rm (#2)}}
\NC{\messagei}        [2]{\messagel{#1}{#2}}
\NC{\messageii}       [3]{\messagel{#1}{#2,#3}}
\NC{\messageiii}      [4]{\messagel{#1}{#2,#3,#4}}
\NC{\messageiv}       [5]{\messagel{#1}{#2,#3,#4,#5}}

\NC{\msgthreshold}    [1]{\messagei{Threshold}{#1}}
\NC{\msgthresholdt}      {\msgthreshold{$\theta$}}
\NC{\msgwhite}        [1]{\messagei{White}{#1}}
\NC{\msgwhiteu}          {\msgwhite{$u$}}
\NC{\msgwhitev}          {\msgwhite{$v$}}
\NC{\msgbounddist}    [3]{\messageiii{BoundaryDistance}{#1}{#2}{#3}}
\NC{\msgbounddistubk}    {\msgbounddist{$u$}{$b$}{$k$}}
\NC{\msgboundid}      [1]{\messagei{BoundaryID}{#1}}
\NC{\msgboundidb}        {\msgboundid{$b$}}

\newcommand{\dhistleftlabel}[1]{\put(-1,-.5){\makebox(1,1)[r]{#1}}}
\newcommand{\alphaylabel}[1]{#1}

\begin{document}
\mainmatter

\title{Neighborhood-Based Topology Recognition in Sensor Networks}

\titlerunning{Topology Recognition in Sensor Nets}  
\author{ S.~P.~Fekete\inst{1}\and A.~Kr{\"o}ller\inst{1}\thanks{Supported by
DFG Focus Program 1126, ``Algorithmic Aspects of Large and Complex Networks'',
Grants Fe 407/9-1 and Fi 605/8-1.} \and D.~Pfisterer\inst{2}\footnotemark[1]
\and S.~Fischer\inst{2} \and C.~Buschmann\inst{2}}
\authorrunning{S.~P.~Fekete et al.}   
\tocauthor{S\'andor P.~Fekete (Braunschweig),
Alexander Kr\"oller (Braunschweig),
Dennis Pfisterer (Braunschweig),
Stefan Fischer (Braunschweig),
Carsten Buschmann (Braunschweig)}
\institute{%
  Department of Mathematical Optimization,\\
  Braunschweig University of Technology,\\
  D-38106 Braunschweig, Germany,\\
  \email{\{s.fekete,a.kroeller\}@tu-bs.de}.\\
\and
  Institute of Operating Systems and Computer Networks,\\
  Braunschweig University of Technology,\\
  D-38106 Braunschweig, Germany,\\
  \email{\{pfisterer,fischer,buschmann\}@ibr.cs.tu-bs.de}.
}

\maketitle              

%
%
%
%
%
%
%
%
%


\begin{abstract} We consider a crucial aspect of self-organization of
a sensor network consisting of a large set of simple sensor nodes with
no location hardware and only very limited communication range. After
having been distributed randomly in a given two-dimensional region,
the nodes are required to develop a sense for the environment, based
on a limited amount of local communication.  We describe algorithmic
approaches for determining the structure of boundary nodes of the
region, and the topology of the region. We also develop methods for
determining the outside boundary, the distance to the closest boundary
for each point, the Voronoi diagram of the different boundaries, and
the geometric thickness of the network.  Our methods rely on a number
of natural assumptions that are present in densely distributed sets of
nodes, and make use of a combination of stochastics, topology, and
geometry.  Evaluation requires only a limited number of simple local
computations.
\end{abstract}




%
%
%
%


\section{Introduction}
\label{sec:intro}

In recent time, the study of wireless sensor networks (WSN) has become
a rapidly developing research area that offers fascinating
perspectives for combining technical progress with new applications of
distributed computing. Typical scenarios involve a large swarm of
small and inexpensive processor nodes, each with limited computing and
communication resources, that are distributed in some geometric
region; communication is performed by wireless radio with limited
range.  As energy consumption is a limiting factor for the lifetime of
a node, communication has to be minimized. Upon start-up, the swarm
forms a decentralized and self-organizing network that surveys the
region.

From an algorithmic point of view, the characteristics of a sensor
network require working under a paradigm that is different from
classical models of computation: Absence of a central control unit,
limited capabilities of nodes, and limited communication between nodes
require developing new algorithmic ideas that combine methods of
distributed computing and network protocols with traditional
centralized network algorithms. In other words: How can we use a
limited amount of strictly local information in order to achieve
distributed knowledge of global network properties?

This task is much simpler if the exact
location of each node is known. Computing node coordinates 
has received a considerable amount of attention.
Unfortunately, computing exact coordinates requires the use of
special location hardware like GPS, or alternatively, 
scanning devices, imposing physical demands on size and structure
of sensor nodes.  
A promising alternative may be 
continuous range modulation for measuring distances between nodes,
but possible results have their limits: The accumulated
inaccuracies from local measurements tend to produce
significant errors when used on a global scale.  This is well-known
from the somewhat similar issue of {\em odometry} from the more
progressed field of robot navigation, where much more powerful
measurement and computing devices are used to maintaining a robot's
location,
requiring additional navigation tools.  
See \cite{stachniss03iros} for some examples and references.
Finally, computation and use of exact coordinates of nodes tends to be
cumbersome, if high accuracy is desired.

It is one of the main objectives of this paper to demonstrate that
there may be a way to sidestep many of the above
difficulties:  Computing
coordinates is not an end in itself. Instead, 
some structural location aspects do {\em not}
depend on coordinates. An additional motivation for our work
is the fact that location awareness for sensor
networks in the presence of obstacles (i.e., in the presence of holes
in the surveyed region) has received only little attention.


One key aspect of location awareness is {\em boundary recognition},
making sensors close to the boundary of the surveyed region 
aware of their position and letting them form
connected {\em boundary strips} along each verge.  
This is of major importance for keeping track of events entering or
leaving the region, as well as for communication purposes to the
outside.  Neglecting the existence of holes in the region may also
cause problems in communication, as routing along shortest paths tends
to put an increased load on nodes along boundaries, exhausting their
energy supply prematurely; thus, a moderately-sized hole (caused by
obstacles, by an event, or by a cluster of failed nodes) may tend to
grow larger and larger.

Beyond discovering closeness to the boundary, 
it is desirable to determine the number and structure of boundaries, 
but also more advanced properties like membership to the 
{\em outer boundary} (which
separates the region from the unbounded portion of the outside world)
as opposed to the {\em inner boundaries} (which separate the swarm
from mere holes in the region). Other important goals are the
recognition of nodes that are well-protected by being far from the
boundary, the recognition of nodes that are on the watershed between
different boundaries (i.e., the Voronoi subdivision of the region),
and the computation of the overall geometric {\em thickness} of the
region, i.e., the size of the largest circle that can be fully
inscribed into the region.

We show that based on a small number of natural
assumptions, a considerable amount of location awareness can indeed be
achieved in a large swarm of sensor nodes, in a relatively simple and
self-organizing manner after deployment, without any use of location
hardware. Our approach combines aspects of random distributions with
natural geometric and topological properties.

\medskip {\bf Related Work.}  There are many papers dealing with node
coordinates; for an overview, consider the cross-references for some
of the following papers.  A number of authors use anchors with known
coordinates for computing node localization, in combination with hop
count.  See
\cite{doherty01convexpositioning,savarese02robust,sundaram02connectivitylocation}.
\cite{capkun01gpsfree} uses only distances between nodes for building
coordinates, based on triangulation.
\cite{priyantha03anchorfreelocalization} presents a fully distributed
algorithm that builds coordinate axes based on a by a near/far-metric
and runs a spring embedder. Many related graph problems are NP-hard,
as shown by \cite{breu93unit} for unit disk graph recognition, and
by \cite{aspnescomputational} for the case of known distances to the
neighbors.

On the other hand, holes in the environment have rarely been considered. 
This is closely related to the $k$-coverage
problem: Decide whether all points in the network area are monitored
by at least $k$ nodes, where $k$ is given and fixed. In this context,
a point is monitored by a node, if it is within the node's sensing
area, which in turn is usually assumed to be a disk of fixed size. By
setting the sensing range to half the communication range,
$1$-coverage becomes a decision problem for the existence of holes. In
\cite{huang03coverage}, an algorithm for $k$-coverage that can be
distributed is proposed. Unfortunately, it requires precise
coordinates for all nodes.  In \cite{fang04hole}, holes are addressed
with greedy geographic routing in mind: Nodes where data packets can
get stuck are identified using a fully local rule, allowing identification
of the adjacent hole by using a
distributed algorithm.  Again, node coordinates must be known for both
detection rule and bypassing algorithm to work.
\cite{aim200317} considers detection of holes resulting from failing
nodes. It proposes a distributed algorithm that uses a hierarchical
clustering to find a set of clusters that touch the failing region and
circumscribe it.

\medskip {\bf Our Results.}  We show that distributed location
awareness can be achieved without the help of location hardware. In
particular:

\begin{itemize}
\item We describe how to recognize the nodes that are near the
  boundary of the region. The underlying geometric idea is quite
  simple, but it requires some effort on both stochastics and
  communication to make it work.
\item We extend our ideas to distinguish the {\em outside boundary}
from the {\em interior boundaries}. 
\item We describe how to compute both boundary distance for all
nodes and overall region thickness.
\item We sketch how to organize communication along the boundaries.
\item We describe how to compute the {\em Voronoi boundaries} that are
halfway between different parts of the boundary.
\end{itemize}

The rest of this paper is organized as follows. In Section~\ref{sec:prelim}
we give some basic notation and state our underlying model assumptions.
In Section~\ref{sec:tree} we describe how to obtain an auxiliary
tree structure that is used for computing and distributing
global network parameters. Section~\ref{sec:prob} gives a brief
overview of probabilistic aspects that are used in the rest
of the paper to allow topology recognition. Section~\ref{sec:bound}
describes how to perform boundary recognition, while Section~\ref{sec:high}
gives a sketch of how to compute more advanced properties.
Section~\ref{sec:experiments} describes implementation issues
and shows some of our experiments. Finally, Section~\ref{sec:future}
discusses the possibilities for further progress based on our work.

\section{Preliminaries}
\label{sec:prelim}

{\bf Swarm and geometry.}
In the following, we assume that the swarm 
consists of a set $V$ of nodes, and the cardinality of $V$ is some
large number $n$. Each node $v\in V$ has a globally unique ID 
(for simplicity, denoted by $v$) of storage size $\bigo{\logn}$,
and coordinates $\posof{v}$ that are unknown to any node.
All node positions are contained in some connected region 
$\netarea\subset\plane$, described by its boundary elements.
In computational geometry, it is common to consider regions
that are non-degenerate polygons, bounded by $k$ disjoint
closed polygonal curves, consisting of a total of $s$ line segments,
meeting pairwise in a total of $s$ corners. Each of the
$k$ boundary curves separates the interior of $\netarea$
from a connected component of $\plane\setminus\netarea$.
The unique boundary curve separating $\netarea$ from 
the infinite component of $\plane\setminus\netarea$ is called
the {\em outside boundary}; all other boundaries are {\em inside boundaries},
separating $\netarea$ from {\em holes}. Thus, the genus of $\netarea$ is
$k-1$.
As we do not care about the exact shapes
(and an explicit description of the boundary is neither required nor
available to the nodes), we do not assume that $\netarea$ is
polygonal, meaning that curved (e.g., circular) instead of linear
boundary pieces are admissible;
we still assume that it consists of $s$ elementary
curves, joined at $s$ corners, with a total number of $k$ boundaries.

\begin{figure}
  \centering
  \setlength{\unitlength}{.5cm}
  \begin{picture}(14,7)
    \put(2,0){\epsfig{file=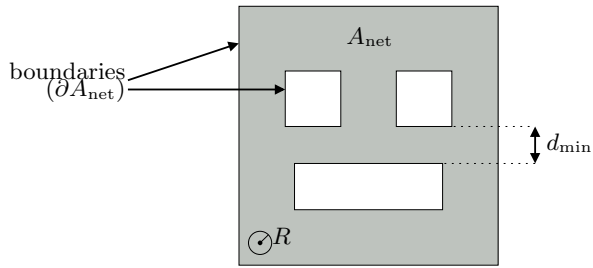,height=7\unitlength}}
    \put(13.2,2.8){\makebox(1,1)[l]{$d_{\min}$}}
    \put(0,4){\makebox(2,1)[rt]{($\netboundary$)}}
    \put(0,5){\makebox(2,.9)[rb]{boundaries}}
    \put(5,5.2){\makebox(7,1.8){$\netarea$}}
    \put(5.9,.6){$\commradius$}
  \end{picture}
  \caption{Geometric parameters.}
  \label{fig:netgeo}
\end{figure}

As narrow bottlenecks in a region can lead to various computational
problems, a standard assumption in computational geometry is to consider
regions with a lower bound on the {\em fatness} of the region; for a polygonal
region, this is defined as the ratio between $d_{\min}$, the smallest distance
between a corner of the region and a nonadjacent boundary segment,
and $d_{\max}$, the diameter of the region.
When dealing with sensor networks, the only relevant parameter for measuring
distances is the communication radius, $\commradius$.
Thus, we we use a similar parameter, called
{\em feature size}, which is the ratio between $d_{\min}$ and 
$\commradius$. For the rest of this paper,
we assume that feature size has a lower bound of 2.
(This technical assumption is not completely necessary, but it simplifies
some matters, which is necessary because of limited space.)
In addition, we assume that angles between adjacent boundary elements 
are bounded away from $0$ and from $2\pi$, implying that there
are no sharp, pointy corners in the region.

\medskip
{\bf Node distribution.}
A natural scenario for the deployment of a sensor network is
to sprinkle a large number of small nodes evenly over a region.
Thus, we assume that the positioning of nodes in the region is
the result of a random process with a uniform distribution 
on $\netarea$. We also assume reasonable density;
in a mathematical sense, this will be made clear further down.
In a practical sense, we assume that each node can communicate
with at least 100 other nodes, and the overall network is connected.

$\vol{\cdot}$ denotes a volume function (\ie, the
Lebesgue measure) on $\plane$, therefore $0<\vol{\netarea}<\infty$.
For simplicity, $\unitvol:=\pi\commradius^2$ denotes the area of the
disk with radius $\commradius$.

Using the notation $\nodesinarea{A}:=\set{\vinV:\posof{v}\in A}$, the expected number
of nodes to fall into an area $A\subset\netarea$ is therefore
\begin{equation}
  \expected{\card{\nodesinarea{A}}}
  =
  n\frac{\vol{A}}{\vol{\netarea}}
  \;.
  \label{eq:generaldensity}
\end{equation}

Therefore, a node $\vinV$ that is not close to the network area's boundary, \ie,
$\ball{\commradius}{\posof{v}}\subset\netarea$ has an
estimated neighborhood size of
\begin{equation}
  \mu := \expected{\card{N(v)}} = (n-1)\frac{\unitvol}{\vol{\netarea}}
  \;.
  \label{eq:density}
\end{equation}
Here, $\ball{r}{x}$ denotes the ball around $x$ with radius $r$.

\medskip
{\bf Node communication.}
Nodes can broadcast messages that are received by all 
nodes within communication range. The cost of broadcasting
one message of size $m$ is assumed to be
$\bigo{m}$; e.g., any message containing a sender ID incurs a
cost of $\bigo{\logn}$.

We assume that two nodes $u\neq\vinV$ can communicate if, and only if,
they are within distance $\commradius$. This is
modeled by a set of edges, \ie,
$uv\inE:\iff\norm{\posof{u}-\posof{v}}\leq\commradius$, where
$\norm{\cdot}$ denotes the Euclidean norm. The set of adjacent nodes
of $\vinV$ is denoted by $N(v)$, and does not include $v$ itself. Such
a graph is known under many names, \eg, geometric, (unit) disk, or
distance graph. The maximum degree is denoted by $\Delta:=\max_{\vinV}\card{N(v)}$.


\section{Leader Election and Tree Construction}
\label{sec:tree}

A first step for self-organizing the swarm of nodes is 
building an auxiliary structure that is used for gathering and
distributing data. The algorithms that are presented in
Section~\ref{sec:bound} only work if certain global network parameters
are known to all nodes. By using a directed spanning tree, nodes know
when the data aggregation phase terminates and subsequent algorithmic
steps may follow. This is in contrast to other methods like flooding,
where termination time is unknown.
The issue of {\em leader election} has been studied in various
contexts; see \cite{brunekreef96design} for a good description.  In
principle, protocols for leader election may be used for our purposes,
as they implicitly construct the desired tree; however, using node IDs
(or pre-assigning leadership) does offer some simplification.

An alternative to leader election is offered by the seminal paper
\cite{gallager83distributedmst} dealing with distributed and emergent
search tree construction. It builds a minimum spanning tree in a graph
with $n$ nodes in a distributed fashion, using only local
communication.  Complexities are $\bigo{n\lognsqr}$ 
time from the first message by a node to completed information at the
constructed root, and $\bigo{\lognsqr}$ transmissions per node,
consisting of $\bigo{\logn}$ messages, each sized $\bigo{\logn}$.

In the following, we will use this auxiliary tree; in particular, we
may assume that each node knows $N(v)$ and $n$, and it is able to use
the tree for requesting and obtaining global data. Note that the tree
is only used for bootstrapping the network; it may be replaced by a
more robust structure at a later time.


\section{Probabilistic Aspects}
\label{sec:prob}

The idea for recognizing boundary nodes is relatively simple: Their
communication range intersects a smaller than average portion of the
region, and thus $N(v)$ is smaller than in other parts. However, a
random distribution of nodes does not imply that the size of $N(v)$ is
an immediate measure for the intersected area, as there may be natural
fluctuations in density that could be misinterpreted as boundary
nodes. In order to allow dealing with this difficulty, we introduce a
number of probabilistic tools.

Recall that \chebychev's inequality shows that for a
binomial distribution for $n$ events with probability $p$, i.e., for
a $\binomial{n}{p}$-distributed random variable~$X$, and
$\alpha<1$
\begin{equation}
  \prob{X\leq\alpha n p}\leq\frac{1}{n}\cdot\mathrm{const} \to 0
  (n\to\infty)
  \label{eq:chebychev}
\end{equation}
holds. We exploit this fact to provide a simple local rule to
let nodes decide whether they are close to the boundary
$\netboundary$. Let $\alpha<1$ be fixed, and let
\begin{equation}
  \boundarynodes =
  \boundarynodes(\alpha) :=
  \set{\vinV: \card{N(v)}\leq\alpha\mu}
  \;.
\end{equation}

\begin{theorem}
  Let $v$ be a node whose communication range lies entirely in
  $\netarea$. Then $v\notin\boundarynodes$ with high probability.
\end{theorem}

\begin{proof}
This follows directly from \eqref{eq:chebychev}, as
\begin{equation}
  \prob{\card{N(v)}\leq \alpha\mu}
  =
  \prob{\card{\nodesinarea{\ball{R}{x_v}}\setminus\set{v}}\leq\alpha\mu}\to 0 \quad(n\to\infty)
  \;.
\end{equation}
\qed
\end{proof}

\begin{theorem}\label{thm:exboundnode}
  Let $x\in\netboundary$ be on the network area's boundary.
  Let $\epsilon>0$. Assume
  $\alpha>\tfrac{1}{\unitvol}{\vol{\ball{R+\epsilon}{x}\cap\netarea}}$.
  Then, with high probability, there is a node $v\in\boundarynodes$
  with $\norm{x-\posof{v}}\leq\epsilon$.
\end{theorem}

\begin{proof}
  \newcommand{\proofarea}{\netball{\epsilon}{x}} Let
  $\proofarea:=\ball{\epsilon}{x}\cap\netarea$ be the area where $v$
  is supposed to be. Then $\vol{\proofarea}>0$ by our assumption on
  feature size.  The probability that there is no node in
  $\proofarea$ equals the probability for a
  $\binomial{n}{\tfrac{\vol{\proofarea}}{\netarea}}$-distributed
  variable to become zero, \ie,
  \begin{equation}
    \prob{\card{\nodesinarea{\proofarea}}=0} =
    \left(1-\tfrac{\vol{\proofarea}}{\netarea}\right)^n
    \to 0 
    \quad (n\to\infty)
    \;.
  \end{equation}
  On the other hand, the probability that a node $u$ in $\proofarea$
  has more than $\alpha\mu$ neighbors is
  \begin{eqnarray*}
    \prob{\card{N(u)}>\alpha\mu}
    &=&
    \prob{\card{\nodesinarea{\proofarea}} > \alpha\mu+1 \given u\mbox{ exists}}
    \\
    &\leq&
    \prob{\card{\nodesinarea{\ball{R+\epsilon}{x}}}>\alpha\mu+1}
    \\
                                %
                                %
    &\to&
    0\quad(n\to\infty)\mbox{, because }\alpha\unitvol>\vol{\netball{R+\epsilon}{x}}
    \;.
  \end{eqnarray*}
  Together, we get 
  \begin{eqnarray*}
    &&
    \prob{\exists\vinV, \posof{v}\in\proofarea: \card{N(v)}\leq\alpha\mu}
    \\
    &=&
    1
    - \prob{\nodesinarea{\proofarea}=\nil}
    - \prob{\forall v\in\nodesinarea{\proofarea}: \card{N(v)}>\alpha\mu \given \nodesinarea{\proofarea}\neq\nil}
    \\
    &\geq&
    1
    - \prob{\nodesinarea{\proofarea}=\nil}
    - \prob{\card{N(v)}>\alpha\mu \given v\in\nodesinarea{\proofarea}}
    \\
    &\to& 1 \quad (n\to\infty)
    \;,
  \end{eqnarray*}
  which proves the claim.\qed
\end{proof}

The assumed lower bound on $\alpha$ can be derived from natural geometric 
properties. For example, if all angles are between $\frac{\pi}{2}$ and 
$\frac{3\pi}{2}$, then for $\alpha> 0.75$ the condition holds for 
a reasonably small $\varepsilon$.
We conclude that $\boundarynodes$ reflects the boundary very closely. It
can be determined by a simple local rule, namely checking whether the
number of neighbors falls below $\alpha\mu$. However, this requires
that all nodes know the value of $\alpha\mu$. 
The next Section~\ref{sec:bound} focuses on this key issue by providing
distributed methods for estimating $\mu$ and $\alpha$.


\section{Boundary Computation}
\label{sec:bound}

\newcommand{\dhistmu}{$\mu$}
\newcommand{\dhistmuest}{$\mu_{\mathrm{est}}$}
\newcommand{\dhistmuhalf}{$\tfrac{1}{2}\mu$}
\newcommand{\dhistmaxdeg}{$\Delta$}
\newcommand{\dhistorigin}{$0$}
\newcommand{\dhistdownlabel}[1]{\put(-1,-1){\makebox(2,1)[t]{#1}}}
\newcommand{\dhistdownleftlabel}[1]{\put(-2,-1){\makebox(2,1)[tr]{#1}}}
\newcommand{\dhistdownrightlabel}[1]{\put(0,-1){\makebox(2,1)[tl]{#1}}}
\newcommand{\alphalabel}[1]{#1}
\newcommand{\alphabordlabel}[1]{{\small #1}}
\newcommand{\dhistuplabel}[1]{\put(-1,.1){\makebox(2,1)[b]{#1}}}

As described in the previous section, the key for deciding boundary
membership is to obtain good estimates for the average density $\mu$
of fully contained nodes, and determining a good threshold $\alpha$.
In the following Subsection~\ref{subsec:mu} we derive a method for
determining a good value for $\mu$.  Subsection~\ref{subsec:algo}
gives an overview of the resulting distributed algorithm, if $\alpha$
has been fixed.  The final Subsection~\ref{subsec:alpha} discusses how
to find a good value for $\alpha$.

\subsection{Determining Unconstrained Average Node Degree $\mu$}
\label{subsec:mu}

Computing the overall average neighborhood size can be performed
easily by using the tree structure described above. However, for
computing $\mu$, we need the average over {\em unconstrained}
neighborhoods; the existence of various pieces of boundary may lower
the average, thus resulting in wrong estimates.

On the other hand, it is not hard to determine the {\em maximum}
neighborhood size $\Delta$. As was shown by
\cite{appel97maximumvertexdegree}, the ratio of maximum to average
degree in the unit square intersection graph of a set of $n$ random
points with uniform distribution inside of a large square tends to 1
as $n$ tends to infinity. We believe that a similar result can be
derived for unit disk intersection graphs.  Unfortunately, convergence
of the ratio is quite slow, and using $\Delta$ as an estimate for
$\mu$ is not a good idea.  For our illustrative example
with 45,000~nodes (see Figure~\ref{fig:standardexample}), 
we get $\Delta/\mu \approx 1.37$.

\begin{figure}
  \centering
  \setlength{\unitlength}{1cm}
  \begin{picture}(9,3.5)
    \put(0,0){
      \input{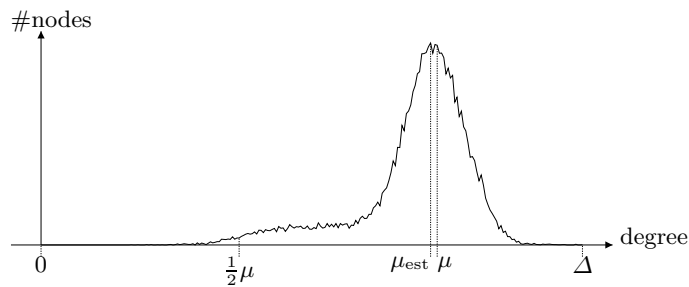}
    }
    \put(0,3){\makebox(2,.5)[bl]{\#nodes}}
    \put(8.1,0){\makebox(.9,.5)[l]{degree}}
  \end{picture}
  \caption{Node degree histogram}
  \label{fig:dhist}
\end{figure}

However, even for very moderate sizes of $n$, $\Delta$ is within a
small constant of $\mu$, allowing us to compute the {\em node degree
  histogram} shown in Figure~\ref{fig:dhist}, again by using the
auxiliary tree structure. 
Clearly, the histogram arises by overlaying three different
distributions:

\begin{enumerate}
\item The neighborhood sizes of all non-boundary nodes.
\item The neighborhood sizes of near-boundary nodes, at varying
distance from the boundary.
\item The neighborhood sizes of boundary nodes.
\end{enumerate}

We expect a pronounced binomial distribution around $\mu$ for (1.), a
uniform distribution for values safely between $\mu/2$ and
$\mu$ for 2., overlayed with a small binomial distribution
for values under $\mu/2$ for 3., possibly skewed in the presence
of many nodes near corners of the region.
(The latter is not to be expected under our geometric
assumption of bounded feature size and minimum angle,
but could be used as an indication of a large number
of pointy corners otherwise.)

Obviously, a variety of other conclusions could be drawn from the node
degree histogram. Here we only use the most common neighborhood size
$\mu_{\mbox{est}}$ as an estimate for $\mu$.  In our example,
$\mu\approx 179.65$ and $\mu_{\mbox{est}}=177$. The according
histogram is shown in Figure~\ref{fig:dhist}. It resembles the
expected shape very closely.

\subsection{Algorithms}
\label{subsec:algo}

When the auxiliary tree is constructed, its root first queries the
tree for $\Delta$, and afterwards for the neighborhood size histogram.
Using $\Delta$, it can quantize the histogram to a fixed number of
entries while expecting a high resolution. This step involves per-node
transmissions of $\bigo{1}$ for queries and $\bigo{\logn}$ for the
responses. On reception of the histogram, the root determines
$\muest$. Assuming that $\alpha$ is known, it then starts a network
flood to pass the value $\alpha\muest$ to all nodes. Message
complexity for the flood is $\bigo{\logn}$.

A node receiving this threshold decides whether it belongs to
$\boundarynodes$. In this case, it informs its neighbors of this
decision after passing on the flood. These nodes form connected
boundaries by constructing a tree as described in
Section~\ref{sec:tree}, with the additional condition that two nodes
in $\boundarynodes$ are considered being connected if their hop
distance is at most 2.

The root of the resulting tree assigns the boundary a unique ID, \eg,
its node ID. This ID is then broadcast over the tree. All nodes
receiving their ID start another network flood. This flood is used
such that each node determines the hop count to its closest boundary.
If it receives messages informing it of two different boundaries at
roughly the same distance, it declares itself to be a Voronoi node.

In addition, the boundary root attempts to establish a one-dimensional
coordinate system in the boundary by sending a message token. The
recipient of this token chooses a successor to forward the token to,
which has to acknowledge this choice. Of the possible successors, \ie,
not explicitly excluded nodes, the one having the smallest common
neighborhood with the current token holder is chosen.  The nodes
receiving the token passing message without being the successor
declare themselves as excluded for futher elections.  After traveling
a few hops, the boundary root gets prioritized in searching for the
token's next hop, thereby closing the token path and forming a closed
loop through the boundary. This path can then be used as axis for the
one-dimensional coordinates.

\subsection{Determining a Good Threshold $\alpha$}
\label{subsec:alpha}

Our algorithms depend on a good choice of the area-dependent parameter
$\alpha$, which should be as low as possible without violating the
lower bound from Theorem~\ref{thm:exboundnode}. If a bound on corner
angles is known in advance, say, $3\pi/2$ in a rectilinear setting,
this is easy: For example, choose $\alpha$ slightly larger than $3/4$.
As this may not always be the case, it is desirable to develop 
methods for the swarm itself to determine a useful $\alpha$.

%

For a too small $\alpha$, no node will be considered part of the
boundary. For increasing $\alpha$, the number of connected boundary
pieces grows rapidly, until $\alpha$ is large enough to allow
different pieces of the same boundary to grow together, eventually
forming the correct set of boundary strips. When further increasing
$\alpha$, additional boundaries appear in low-density areas,
increasing the number of identified boundaries. These boundaries also
begin to merge, until eventually a single boundary consisting of the
whole network is left.


\begin{figure}
  \centering
  \setlength{\unitlength}{1cm}
  \begin{picture}(9,3.5)
    \put(0,0){
      \input{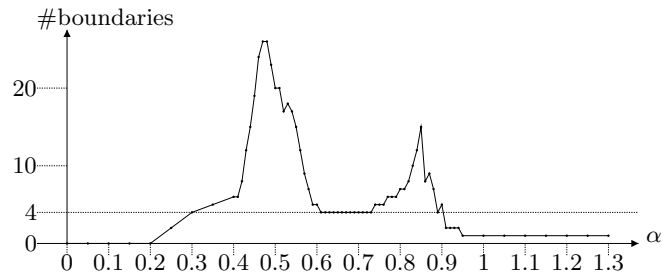}
    }
    \put(0,3){\makebox(2,.5)[bl]{\#boundaries}}
    \put(8.1,0){\makebox(.9,.5)[l]{$\alpha$}}
  \end{picture}
  \caption{The number of boundary components as a function of $\alpha$.}
  \label{fig:ahist}
\end{figure}

Figure~\ref{fig:ahist} shows that this expected behavior does indeed
occur in reality: Notice the clear plateau at 4 connected components,
embedded between two pronounced peaks. 

This shows that computing a good threshold can be achieved by sampling possible
values of $\alpha$ and keeping track of the number of connected boundary
components.


\section{Higher-Order Parameters}
\label{sec:high}

Once the network has identified boundary structures, it is possible
to make use of this structure for obtaining higher-order information.
In this section, we sketch how some of them can be determined.

\subsection{Detection of Outer Boundary}
\label{subsec:outer}
\begin{figure}[htbp]
  \centering
  \newlength{\subfsize}
  \setlength{\subfsize}{.30\linewidth}
  \subfigure[The area of outside and inside strips.]{
    \epsfig{figure=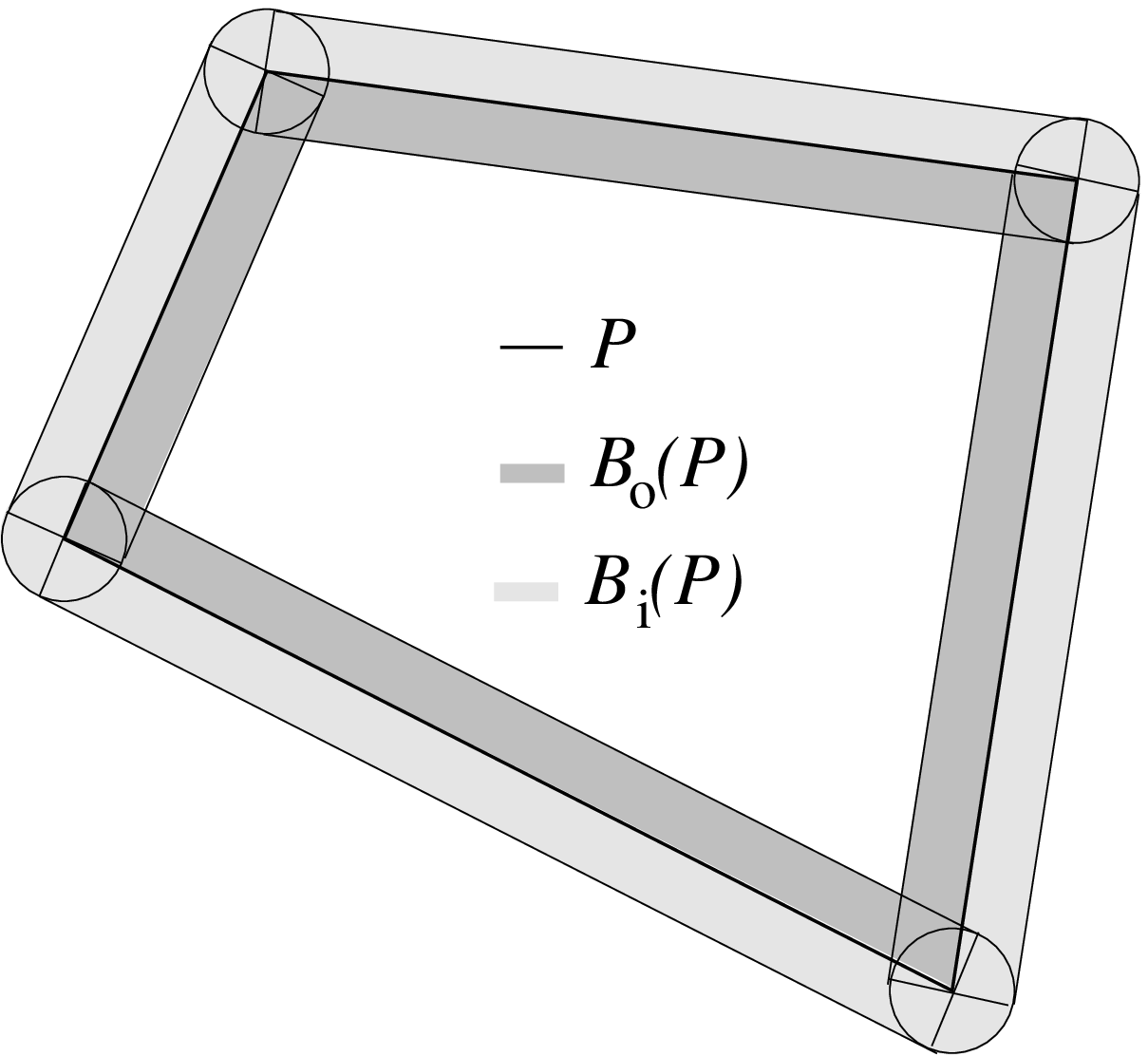,width=\subfsize}
    }
  \hspace{.1\linewidth}
  \subfigure[Overlap between adjacent strips.]{
    \epsfig{figure=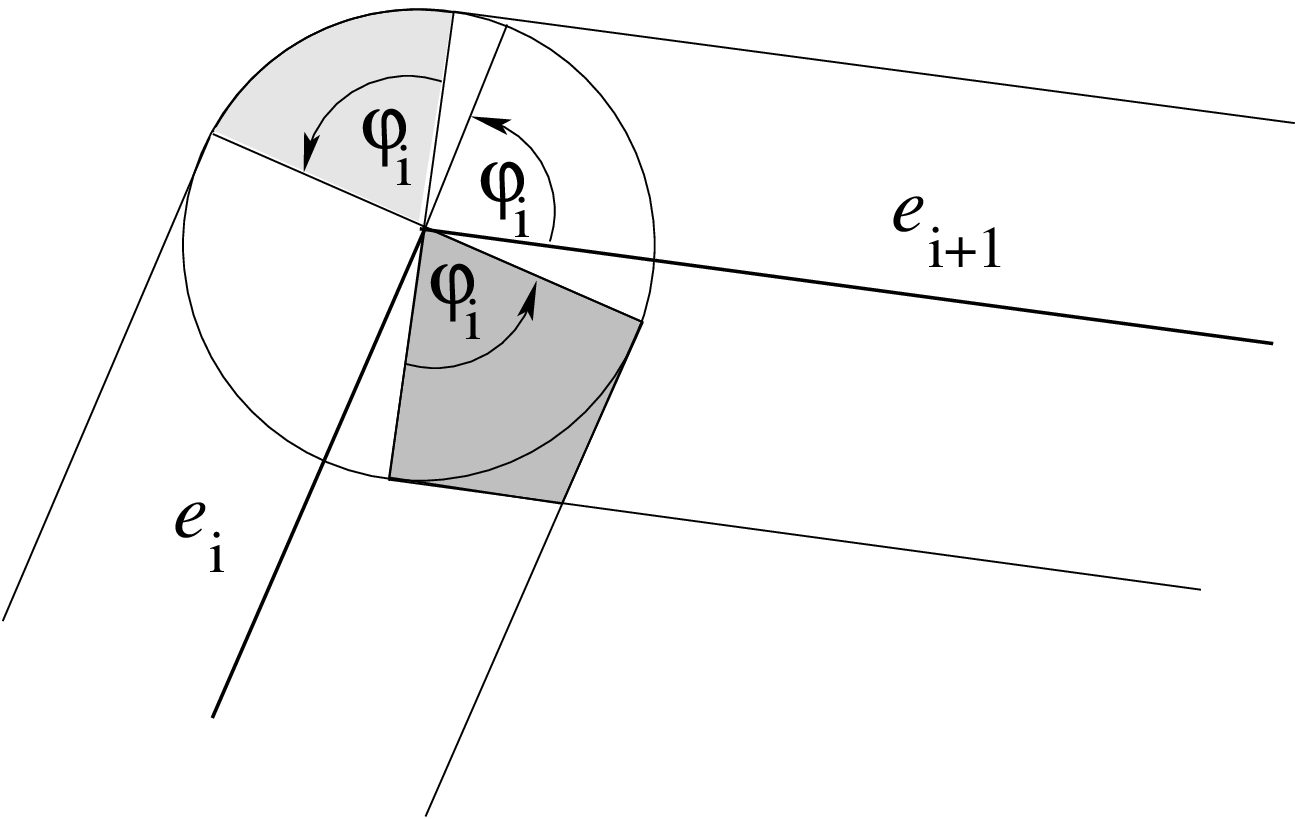,width=\subfsize}
  }
  \vspace{-2ex}
  \caption{\label{fi:strip}
    Geometry of boundary strips.}
\end{figure}

One possible way to guess the outer boundary is to hope that the shape
of boundary curves is not too complicated, which implies that the
outside boundary is longest, and thus has the largest number of
points.  An alternative heuristic is motivated by the following
theorem.  See Figure~\ref{fi:strip} for the idea.


\begin{theorem}
\label{th:strip}
Let $P$ be a simple closed polygonal curve with feature size at least
$2\commradius$ and total Euclidean length $\ell(P)$, consisting of
edges $e_i$, and let $\varphi_i$ be the (outside) angle between edges
$e_i$ and $e_{i+1}$.  Let $B_i(P)$ be the set of all points that are
near $P$ as an inner boundary, i.e., that are outside of $P$ and
within distance $\commradius$ of $P$, and let $B_o(P)$ be the set of
all points that are near $P$ as an outside boundary, i.e., that are
inside of $P$ and within distance $\commradius$ of $P$.  Then the area
of $B_o(P)$ is
$\commradius\ell(P)-\sum_{\varphi_i>0}\commradius^2\frac{\varphi_i}{2}+
\sum_{\varphi_i<0}\commradius^2\tan(\frac{-\varphi_i}{2})
\pi\commradius^2$, while the area of $B_i(P)$ is
$\commradius\ell(P)+\sum_{\varphi_i<0}\commradius^2\frac{\varphi_i}{2}-
\sum_{\varphi_i>0}\commradius^2\tan(\frac{-\varphi_i}{2})
\pi\commradius^2$.
\end{theorem}

\begin{proof}
  As shown in Figure~\ref{fi:strip}, both $B_o(P)$ and $B_i(P)$ can be
  subdivided into a number of strips $s_i$ parallel to edges $e_i$ of
  $P$, and circular segments near vertices of $P$ that are either
  positive (when the angle between adjacent strips is positive,
  meaning there is a gap between the strips) or negative (when the
  angle is negative, meaning that strips overlap.) More precisely, if
  the angle $\varphi_i$ between edges $e_i$ and $e_{i+1}$ is positive,
  we get an additional area of $\commradius^2\frac{\varphi_i}{2}$,
  while for a negative $\varphi_i$, the overlap is
  $\commradius^2\tan(\frac{\varphi_i}{2})$. In total, this yields the
  claimed area. Our assumption on feature size guarantees that no
  further overlap occurs.
  \qed
\end{proof}

Note that in any case, $\sum_i{\varphi_i}=2\pi$.  Making use of this
property is possible in various ways: It is natural to assume that the
area for the exterior boundary strip is less than suggested by its
length, while it should be larger for all other boundaries. This
remains true for other kinds of boundary curves using similar
arguments.

A straightforward estimate for strip area is given by the number
$|N(D_j)|$, while $|D_j|$ is a natural estimate for the length of the
boundary, as the density of boundary nodes should be reasonably
uniform along the boundary.  Near boundary corners, the actual number
of boundary nodes will be higher for convex corners (as the threshold
of neighborhood size remains valid at larger distance from the
boundary), and lower for nonconvex corners.  Thus it makes sense to
consider the ratio $\frac{|N(D_j)|}{|D_j|}$ for all boundary
components, as the outside boundary can be expected to have lower than
average $|N(D_j)|$ and higher than average $|D_j|$.  The component
with the lowest such ratio is the most likely candidate for being the
outside boundary.  See Table~\ref{tab:strip} for the values of our
standard example.

\begin{table}[htbp]
\begin{center}
\begin{tabular}{|c||c|c|c|c|}
\hline
            & $D_1$ & $D_2$ & $D_3$ & $D_4$ \\
            & (Outside boundary) & (Left eye) & (Right eye) & (Mouth) \\\hline\hline 
$|N(D_j)|$ &   6093 &   1304 &   1319 & 2368 \\\hline
$|D_j|$ &    2169 & 289 & 266 & 616 \\\hline
$\frac{|N(D_j)|}{|D_j|}$ &    2.809 & 4.512 & 4.959 & 3.844 \\\hline
\end{tabular}
\par\medskip
\caption{\label{tab:strip}Number of boundary nodes and near-boundary nodes
for each boundary component.}
\end{center}
\end{table}

The main appeal of this approach is that the required data is already
available, so evaluation is extremely simple; we do not even have to
determine hop count along the boundary.

It should be noted that our heuristic may produce wrong results if
there is an extremely complicated inside boundary.  This can be fixed
by keeping track of angles (or curvature) along the boundary; however,
the resulting protocols become more complicated, and we leave this
extension to future work.

\subsection{Using Boundary Distance}
\label{subsec:distance}

Once all boundaries have been determined, it is easy to compute {\em
  boundary distances} for each node by determining a hop count from
the boundary. Note that this can be done to yield non-integral
distances by assigning fractional distances to the near-boundary nodes,
depending on their neighborhood size.

This makes it easy to compute the geometric {\em thickness} of the
region: Compute a node with maximum boundary distance. In our standard
example, this node is located between the three inside boundaries.


\section{Experimental Results}
\label{sec:experiments}

\begin{figure}
  \setlength{\subsize}{.45\textwidth}
  \subfigure[Boundary nodes, near-boundary nodes, and Voronoi nodes.]{
    \epsfig{file=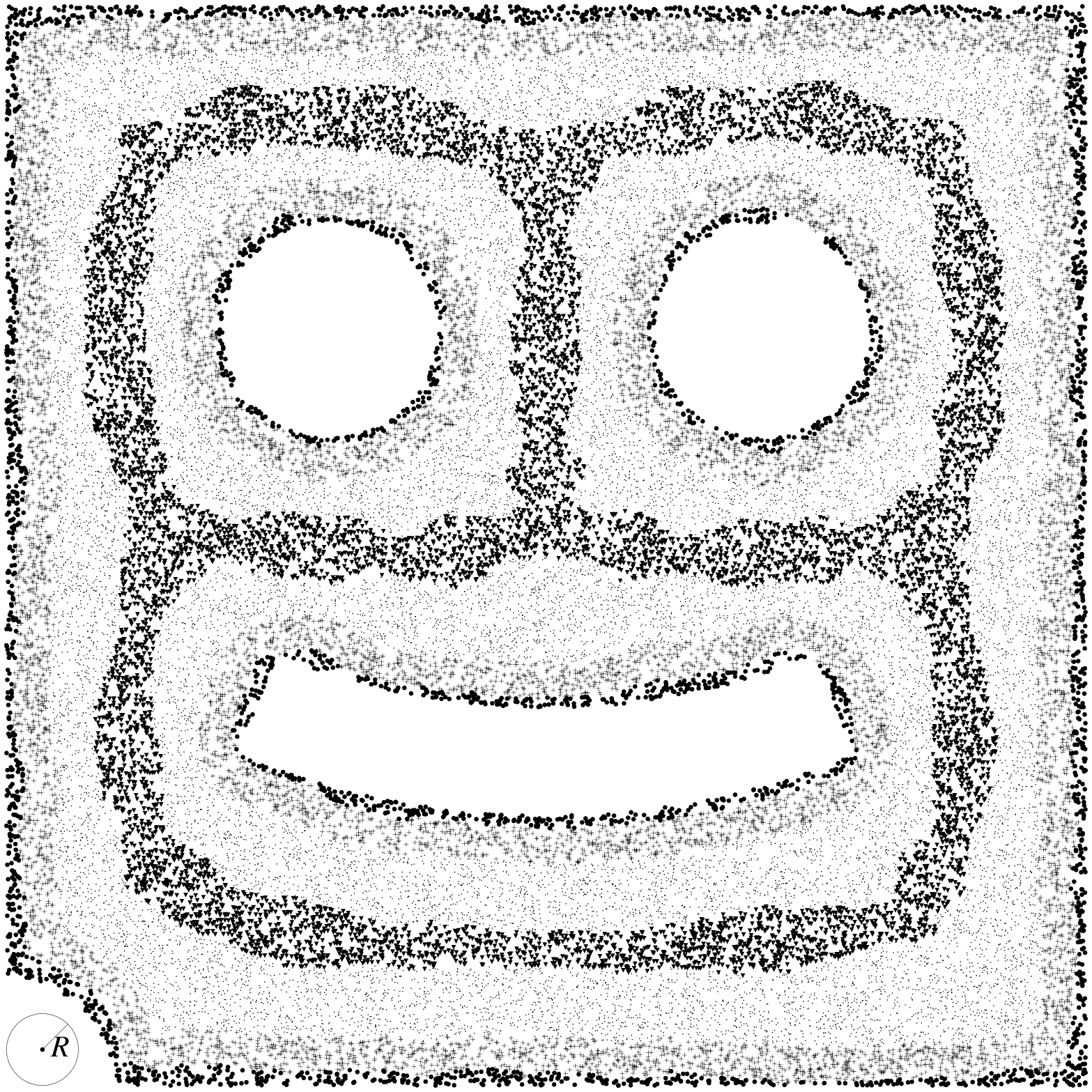,width=\subsize,height=\subsize}
    \label{fig:bound}
  }
  \hfill
  \subfigure[Structural loops along the boundaries.]{
    \epsfig{file=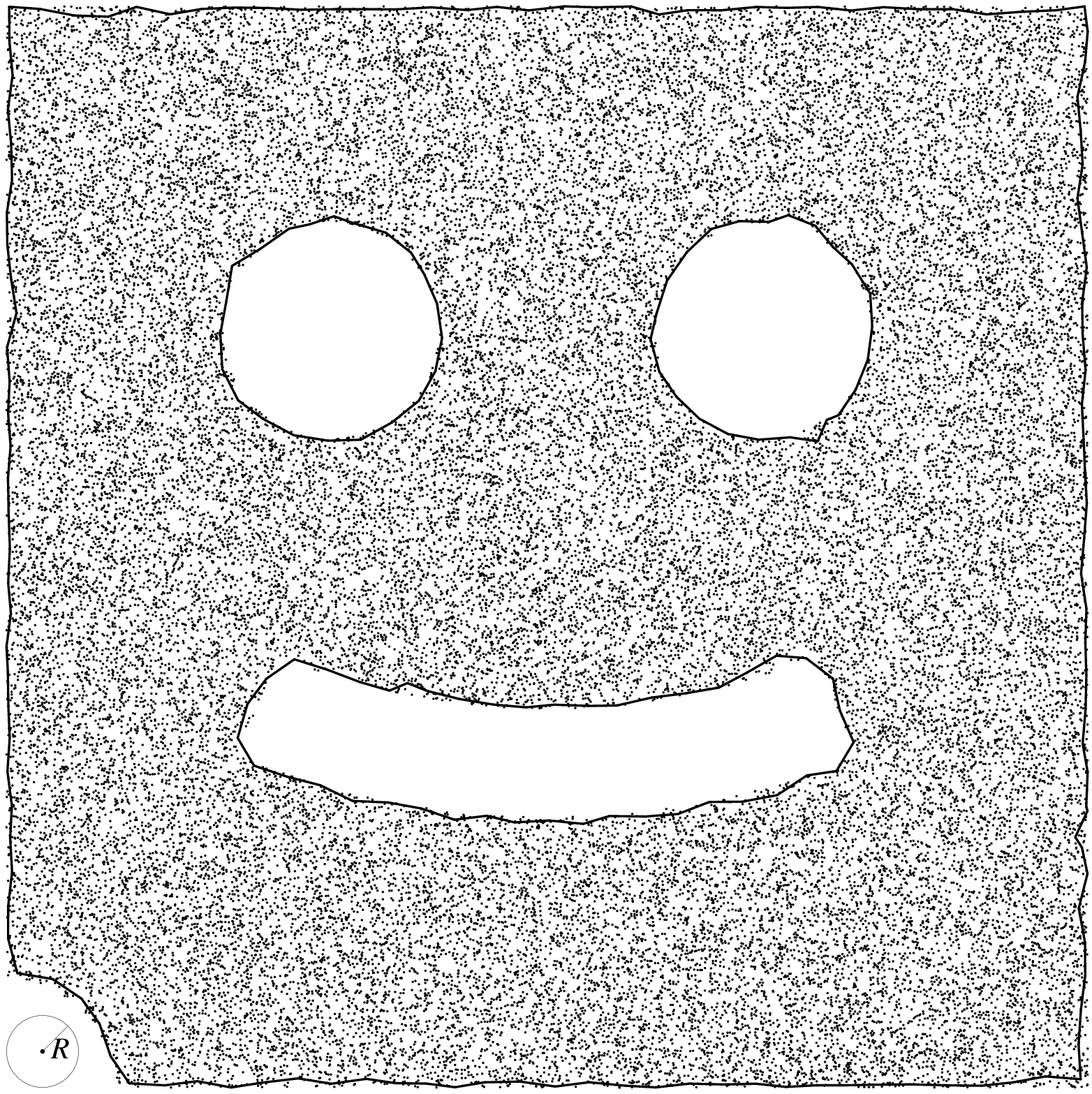,width=\subsize,height=\subsize}
    \label{fig:connect}
  }
  \vspace{-3ex}
  \caption{Experimental results for the example network.}
\end{figure}

All our above algorithms have been implemented and tested on
different point sets. See the following Figure~\ref{fig:topo}
for an example with 45,000 nodes and four boundaries. 
The bounding box has a size of $30\commradius\times 30\commradius$.
Total area of the region is $786.9\commradius^2$. 
Figure~\ref{fig:3D} shows the spatial distribution of neighborhood size.
Notice the slope near the boundaries. 
Figure~\ref{fig:bound} shows the identified boundary, near-boundary
and Voronoi nodes, shown as black dots, gray crosses, black triangles,
while other interior nodes are drawn as thin gray dots. The total
number of identified boundary or near-boundary nodes is 11,358, leaving
33,642 nodes as interior nodes.
Finally, Figure~\ref{fig:connect} shows the assigned structural 
loops along the various boundary strips.

\begin{figure}
  \newlength{\subsize}
  \setlength{\subsize}{.45\textwidth}
  \subfigure[Sensor network topology.]{
    \epsfig{file=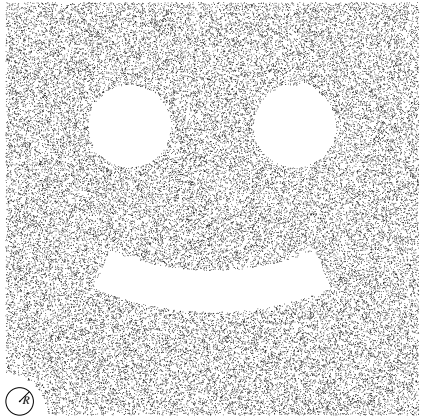,width=\subsize,height=\subsize}
    \label{fig:topo}
    }
  \hfill
  \subfigure[Spatial distribution of $\card{N(v)}$.]{
    \epsfig{file=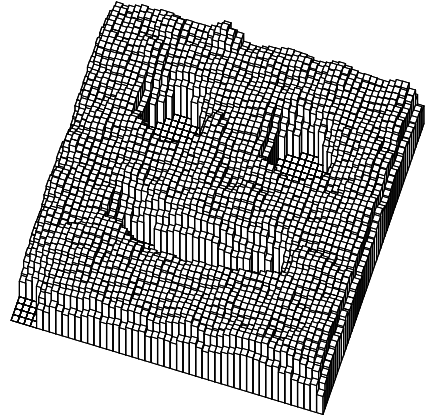,width=\subsize,height=\subsize}
    \label{fig:3D}
  }
  \vspace{-3ex}
  \caption{Example network consisting of 45,000 nodes.}
  \label{fig:standardexample}
\end{figure}

\section{Conclusions}
\label{sec:future}

We have shown that dealing with topology issues in a 
large and dense sensor network is possible, even in the absence of
location hardware or the computation of coordinates.
We hope to continue this first study in various ways.
One possible extension arises from recognizing more detailed
Voronoi structures by making use of the shape of the 
{\em boundary distance terrain}: This shape differs for
nodes that have only one closest line segment in the boundary,
as opposed to nodes that are close to two different such segment,
constituting a ridge in the terrain. Note that our Voronoi nodes
are close to pieces from two {\em different} boundaries.

An obvious limitation of our present approach is the requirement
for high density of nodes.
A promising avenue for overcoming this deficiency is to exploit 
higher-order information
of the neighborhood structure, using more sophisticated
geometric properties and algorithms. This should also allow
the discovery and construction of more complex aspects of the network,
e.g., for routing and energy management.

%
%
%
%
%


\section*{Acknowledgments}
\label{sec:acks}

We thank Martin Lorek for his help.

%
%
% ...
%
%
%
%


\bibliographystyle{alpha}
\bibliography{references}

\end{document}